\documentclass[letterpaper]{article}

\usepackage{natbib,alifeconf}  
\usepackage{amsmath}
\usepackage{amssymb}
\usepackage{caption}
\usepackage{subcaption}
\usepackage[T1]{fontenc}
\usepackage{url}
\usepackage{CJKutf8} 

\title{ALIFE2023 template}

\title{Exploring the Adaptive Behaviors of Particle Lenia: \\A Perturbation-Response Analysis for Computational Agency}
\author{Kazuya Horibe$^1$, Keisuke Suzuki$^2$, Takato Horii$^1$ \and Hiroshi Ishiguro$^1$ \\
$^1$Osaka University $^2$Hokkaido University \\
horibe.289@gmail.com} 

\begin{document}
\begin{CJK}{UTF8}{min}
\maketitle

\begin{abstract}
A firm cognitive subject or ``individual'' is presupposed for the emergence of mind. However, with the development of recent information technology, the ``individual'' has become more dispersed in society and the cognitive subject has become increasingly unstable and adaptive, necessitating an update in our understanding of the individual''. Autopoiesis serves as a model of the cognitive subject, which is unstable and requires effort to maintain itself to adapt to the environment. In this study, we evaluated adaptivity for a highly extensible multi-particle system model Particle Lenia through the response perturbation. As a result, we found that Particle Lenia has a particle configuration that is both temporally unstable and has multiple stable states. This result suggests that Particle Lenia can express adaptive characteristics and is expected to be used as a computational model toward building an autopoietic cognitive agent.
\end{abstract}

\section{Introduction}

The concept of autopoiesis, as proposed by  Maturana and Varela \citep{varela1974organization, maturana1980autopoiesis}, posits that the mind is comprised of emerging autonomous processes that distinguish the system itself from its environment. Autopoiesis denotes an innate attribute of living beings where the boundary between the system and its surroundings is not initially well-established, although rather determined and preserved by interactions with the environment. The notion of autopoiesis has been applied in a multitude of research such as artificial life, biology, cognitive science, and robotics, and has continued to be expanded to this day \citep{razeto2012autopoiesis}. Autopoiesis has been expanded as a framework to explain the origin of cognition in living organisms, leading to the development of the enaction approach \citep{varela1991embodied, thompson2007mind}. The enaction concept encompasses autonomy, operational closure, precariousness, adaptivity, agency, and sense-making as its core features \citep{di2005autopoiesis, di2009extended, di2014enactive, thompson2007mind}.

Computer simulations for autopoietic and enaction features have been conducted to determine the feasibility of autopoiesis \citep{mcmullin2004thirty}. Studies aimed at computational autopoiesis originated from the SCL model, an artificial chemistry model developed by Mcmullin and Varela \citep{mcmullin1997rediscovering}. Suzuki and Ikegami expanded the SCL model to include movement and the ability to maintain its own membrane while in motion \citep{Suzuki2009Shapes, ikegami2008homeostatic}. Beer developed an autopoietic/enactive theory based on the ``glider'' movement pattern, a well-known pattern found in the Game of Life cellular automaton, as a cognitive agent \citep{beer2004autopoiesis, beer2014cognitive, beer2020investigation, beer2023theoretical}. Due to the difficulty of describing the complexity of cognition in cellular automaton-based models, various models based on category theory \citep{letelier2003autopoietic}, finite-state automata \citep{carter2018emergence}, artificial neural networks \citep{froese2023autopoiesis}, and multi-particle systems \citep{friston2013life} have been developed to describe the internal state of a cognitive agent and its hierarchical structure.

This study investigates the autopoietic property especially adaptivity using a Particle Lenia model, which is a highly expressive multi-particle system model \citep{alexander2023particle} to investigate a scalable autopoietic/enation system. Particle Lenia, an extension of the expressive multi-particle system model Lenia \citep{chan2018lenia}, explores a diverse range of patterns resembling over 400 biological creatures using differentiable computation within a scalable continuous cellular automaton framework. Particle Lenia gains the flexibility to adapt different rules and parameters to each particle and easily extend the three-dimensional space. In this paper, we take the first step in investigating the basic properties of Particle Lenia's adaptivity by measuring the response to perturbations in particle configuration from a stable state.

\section{Related works}
\subsection{Agency through interaction between environment}
Numerous techniques have been proposed for modeling cognitive agents, almost all of which highlight the importance of the agent's interactions with its environment. However, each approach offers a distinct perspective on how the agent engages with its environment and how such interaction influences its cognitive processes. The autopoiesis approach underscores the significance of the agent's internal structure, which enables it to generate self-organized behavior and meaning via its environmental interactions \citep{varela1974organization}. The enactive approach, on the other hand, highlights the importance of the agent's sensory-motor interactions in cognition, suggesting that its behavior arises from its continuous interaction with the environment \citep{di2014enactive}. Embodied cognition stresses the importance of the agent's physical embodiment and its interaction with the environment in shaping its cognitive processes \citep{pfeifer2006body, tani2016exploring, taniguchi2023world}. Finally, the subsumption architecture approach highlights the direct action of the agent in the environment, without the need for internal representations, and proposes a layered control architecture for this purpose \citep{braitenberg1986vehicles, brooks1986robust}. Each of these approaches offers a unique advantage in comprehending different aspects of an agent's cognition. To fully comprehend an agent's cognition, it is important to integrate these approaches and develop a comprehensive understanding of how the agent interacts with its environment.

\subsection{Lenia: cellular automaton-based model}
Lenia is a model of a continuous cellular automaton that has found similar patterns in diverse organisms \citep{chan2018lenia}. Lenia is characterized by the flexibility in defining the interaction cells and updating rules. Previously, cellular automata patterns were commonly broken when noise was added \citep{neumann1966self}, whereas by extending Lenia's state update law to multiple channels, it is now possible to express patterns that are robust against noise \citep{chan2020lenia}. In addition, by differentiating the computation, scalable computation is available on GPUs, and in combination with evolutionary computation, it is possible to automatically search for a pattern that is suitable for a specific purpose, such as having the ability to move. Furthermore, by defining one of the channels as an ``obstacle'' channel, which is an area that the pattern cannot penetrate, the pattern can be extended to have a sensori-moter loop function, in which the agent detects an ``obstacle'' and changes its movement direction \citep{hamon:hal-03519319}. 

Lenia is a cellular automaton-based model that applies the same rules to all of space for computation, making it difficult to interact with patterns generated by different rules. To solve this problem, Flow Lenia has been developed to apply local mass conservation laws to cellular automata, enabling a variety of patterns generated by different rules to coexist in the same space \citep{plantec2022flow}. Particle Lenia used in this study was proposed at the same time as another solution to the problem solved by Flow Lenia \citep{alexander2023particle}.

\subsection{Multi particle system for artificial life}
There have been many models of multi-particle systems proposed for artificial life and self-organizing phenomena. The Boid model is the well-known multi-particle system model that describes flocks of birds and animals with simple rules and can represent a variety of behaviors of flocks \citep{reynolds1987flocks}. The swarm chemistry is an extension of the boid model, in which each particle has its own set of parameters such as the parameters of the boid model, and by adjusting the ratio of the number of particles with the same parameters and mixing them together, a variety of structures can emerge \citep{sayama2009swarm}. The swarm oscillator represents particles as oscillators, each of which has a unique frequency and interacts with each other, and it can express synchronous phenomena that were difficult to express using only the parameters of the boid model \citep{tanaka2007general,ceron2023diverse}. Friston introduced the free energy principle to many-particle systems and proposed an energy-based particle system \citep{friston2013life}. Minimizing the free energy can be interpreted as building a predictive model of the environment, which enables us to describe the motivation of the particle agent. The predictive model of the particle agent is defined as a Markov blanket of a Bayesian network based on physical particle configurations. Particle Lenia, as well as Lenia, can describe interactions with higher degrees of freedom due to the functional type degrees of freedom of the kernel and growth function and their multi-layering and also can analyze free energy due to the energy-based particle system.

\section{Method}
\subsection{Model}
To evaluate the adaptivity, we used Particle Lenia. Particle Lenia \citep{alexander2023particle} is a multi-particle model that rewrites a continuous cellular automaton model called Lenia \citep{chan2018lenia}, which can express diverse patterns. 

Each particle satisfies the following differential equation:
\begin{equation}
    \frac{d{\bf p}_i}{dt} = - \nabla {\bf E}({\bf p}_i) = -  \lbrack \frac{\partial {\bf E}({\bf p}_i) }{\partial {\bf p}_i} \rbrack ^\mathsf{T} ,
\end{equation}
where $\bf {p_i}$ is the position of each particle. The ${\bf E}$ represents the energy of the system $ \rm {\bf{E}  = \bf{R - G}}$, where ${\bf R}$ and ${\bf G}$ can be written, respectively, as follows.

\begin{align}
   {\bf R}^t({\bf x}) & = \frac{{\rm c_{rep}}}{2} \sum_{i:{\bf p}_i \neq {\bf x}} { \max(1-||{\bf x}-{\bf p}_i||, 0)^2}, 
\end{align}
where, $\rm {c_{rep}}$ defines the repulsion strength. ${\bf R}^t$ is called repulsion potential field, where the distance is less than 1, the distance is multiplied by $\rm {c_{rep}}$, and all combinations are combined. Namely, particles with a distance of 1 or more are not repulsive.

${\bf G}^t$ called growth field is represented by ${\bf U}^t$ -- Lenia filed and ${\bf K}^t$ -- kernel, as follows.

\begin{equation}
    \begin{aligned}
    {\bf G}^t({\bf x}) & = G({\bf U}^t({\bf x})) \\
    & = G(\sum_i {\bf K}({\bf x} - {\bf p}^t_i) ) \\
    & = G(\sum_i K(|| {\bf x} - {\bf p}^t_i||) ),
    \end{aligned}
\end{equation}

where $G(u) = \exp(-(u- \mu_G)^2/\sigma^2_G)$ and  $K(r) = w_K \exp(-(r- \mu_K)^2/ \sigma_K^2)$; exponential functions are employed for this paper of $K$ and $G$. $K$ corresponds to the adjacent cell rule in the game of life and determines which adjacent particle is used for the next update of its own state; $G$ corresponds to the growth function in the game of life and determines its next state. ${\mu}_*$ and ${\sigma}_*$ in $K$ and $G$ weight the states of other particles for use in updating its own state. $w_K$ is chosen so that the integral of $K$ over the whole space equals one. In this paper, we used these parameter sets:$\mu_K = 4.0, \sigma_K = 1.0, w_K = 0.022, {\rm c_{rep}}=1.0$. All numerical simulations were performed in the 4th Runge-Kutta method with $dt=0.01$.

\subsection{Measurement for adaptivity}
Adaptivity is a system's ability to adjust its internal states and interactions with the environment in order to maintain viability \citep{di2005autopoiesis}. It involves recognizing and responding to tendencies that bring it closer to or farther from the edge of viability, transforming harmful tendencies into beneficial ones. To characterize the adaptivity in Particle Lenia, we defined two measures; the diversity of metastable configuration and time instability. The diversity of metastable configurations allows multiple configurations to be realized and respond in various ways to perturbations. High-time instability enables the system to reach another stable state by multiple pathways and has a high adaptive capacity to perturbations.

\begin{figure}[ht]
      \begin{minipage}[b]{1\hsize}
        \centering
        \includegraphics[width=1\textwidth]{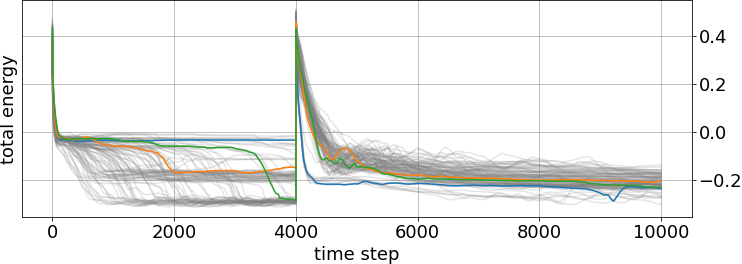}
        \subcaption{Time evolution of total energy}
      \end{minipage} \\
      \begin{minipage}[b]{1\hsize}
        \centering
        \includegraphics[width=1\textwidth]{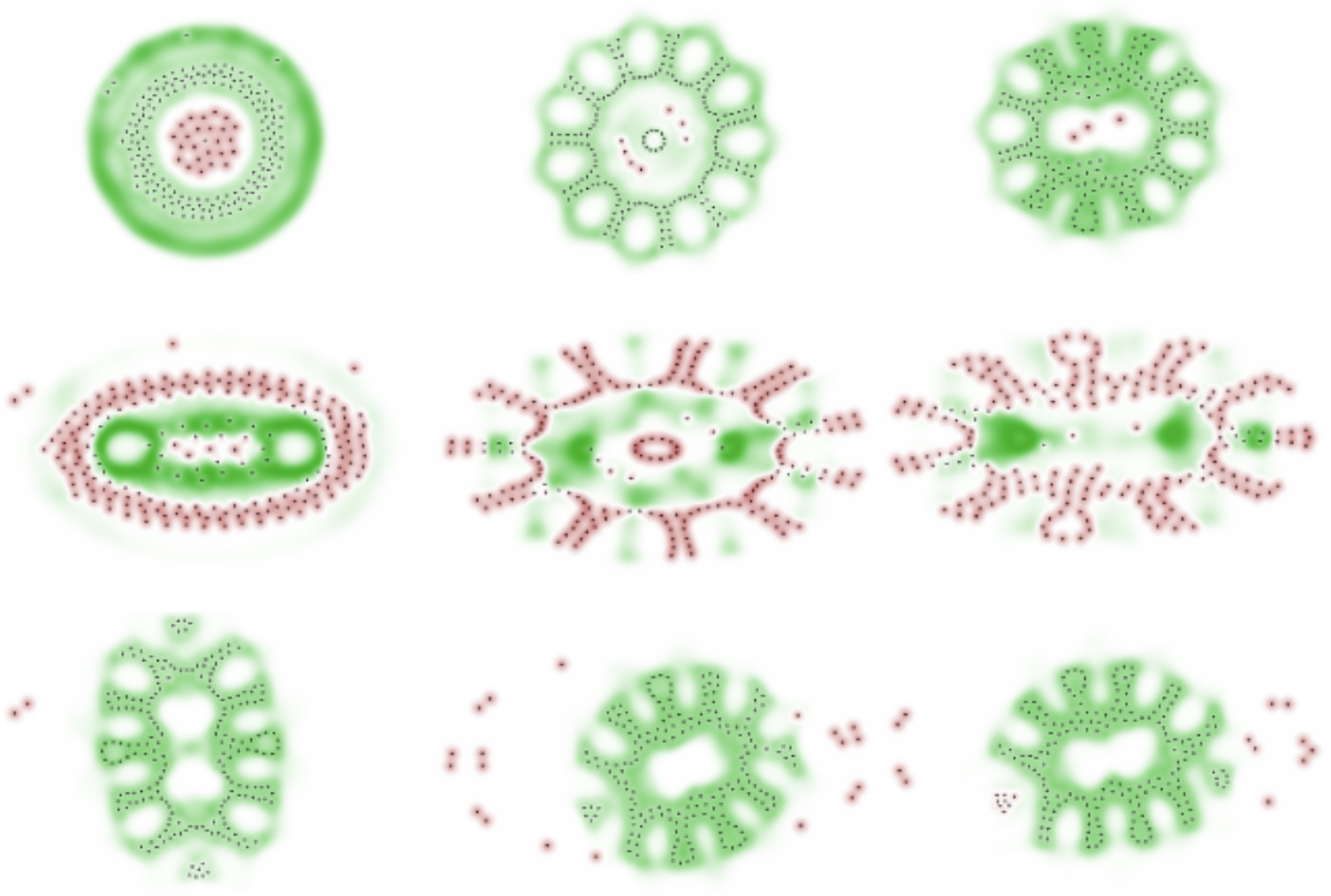}
        \subcaption{Example of particle configuration}
      \end{minipage} 
      \caption{Response to ``stretch'' perturbation. (a) Gray lines show 100 trials. Three of them are selected and displayed in blue, orange, and green lines. Each line corresponds to the left, middle, and right in (b). (b) Upper panel: stable configuration before perturbation (time step = 4,000), middle panel(time step = 4,001): just after perturbation, lower panel: stable configuration after perturbation (time step = 10,000). Green and red are low and high-energy, respectively.}
      \label{fig:fig1}
\end{figure}

The diversity of metastable configurations is a measure of adaptivity to space, which quantifies the variety of possible metastable particle configurations that can be achieved, each with distinct energy values. In each simulation, we recorded the total energy in the time direction over 2,000 steps toward the end of the calculation (specifically, from 8,000 to 10,000 steps). We then calculated the average total energy for each trial and determined the standard deviation of these averages across all 100 trials (Figure \ref{fig:fig1}(a)).

Time instability is a measure of adaptivity to time, which assesses ``how much the energy value fluctuates before reaching a stable particle state.''  In each trial, we monitored the total energy fluctuations in the time direction over 2,000 steps toward the end of the calculation (specifically, from 8,000 to 10,000 steps). We then computed the standard deviation of the total energy for each trial to capture the extent of energy fluctuations during the simulation. Finally, we averaged these standard deviations across all 100 trials to obtain the time instability metric (Figure \ref{fig:fig1}(a)).

\section{Results}
\subsection{Response to ``stretch'' perturbation}
\label{sec:result1}

To investigate the adaptivity of Particle Lenia, perturbations that do not change the number of particles, such as ``stretch'' perturbations, were given and their statistical properties were investigated. For the ``stretch'' perturbation, the x-axis of the particle position was extended in the x-axis direction by applying the linear transformation $x_{new} = 2*x$, and the numerical calculation was resumed from that arrangement. There were $N=200$ particles and the initial placement was sampled according to a uniform distribution from spatial coordinates of $ -6< x < 6, -6< y < 6$. We ran 100 independent simulations with different random seeds of uniform distribution. 

Figure \ref{fig:fig1}(a) shows the time evolution of the total energy for 100 independent simulations. The total energy ${\rm E_{total}}$ was obtained by linearly summing the elements of ${\bf E}$ and dividing by the number of particles. We performed calculations up to 4,000 steps, at which point we applied a ``stretch'' perturbation to the $x$ direction.We restarted the numerical calculations from the stretched particle and computed up to 10,000 steps. Before the perturbation, three `stable' states with different energies were observed around ${\rm E_{total}} = 0, -0.2,$ and $-0.3$. As soon as the ``stretch'' perturbation was applied, the energy of the whole system increased. The energy then decreased rapidly and converged to close energy values for all 100 trials.

We confirmed the configuration of the particles before and after the ``stretch'' perturbation. We found that the stable states with different energies before the perturbation also have different particle configurations. The stable state near ${\rm E_{total}} = 0$ (blue line in Figure \ref{fig:fig1}(a)) is circular, with a high-energy particle in the center surrounded by low-energy particles (Figure \ref{fig:fig1}(b) upper panel, left). There is a ring-shaped band with lower energy further out, but only a few particles were present there. The stable state around ${\rm E_{total}} = -0.2$ (orange line in Figure \ref{fig:fig1}(a)) had a ring of low-energy particles in the center, with a few sparse high-energy particles just outside, and further outside, low-energy particles forming a compartment (Figure \ref{fig:fig1}(b) upper panel, middle). The stable state around ${\rm E_{total}} = -0.3$ (green line in Figure \ref{fig:fig1}(a)) had few particles in the center of the pattern and a few high-energy particles. It was surrounded by a thick band of low-energy particles. Towards the outside of the band, the low-energy particles were spiked, resulting in an overall flattened arrangement (Figure \ref{fig:fig1}(b) upper panel, right). Immediately after the perturbation, the outer particles with lower energies in their respective stable states had higher energies, and ${\rm E_{total}}$ showed higher values. (Figures \ref{fig:fig1}(a) and (b) middle panel). After perturbation, the particles converged to a stable ``rotor'' with decreasing total energy (Figures \ref{fig:fig1}(a) and (b) lower panel). In summary, we confirmed different stable particle configurations with different energies before perturbation, with particles converging to a stable ``rotor'' with decreasing total energy after perturbation.

\begin{figure}[ht] 
      \centering
      \includegraphics[width=.48\textwidth]{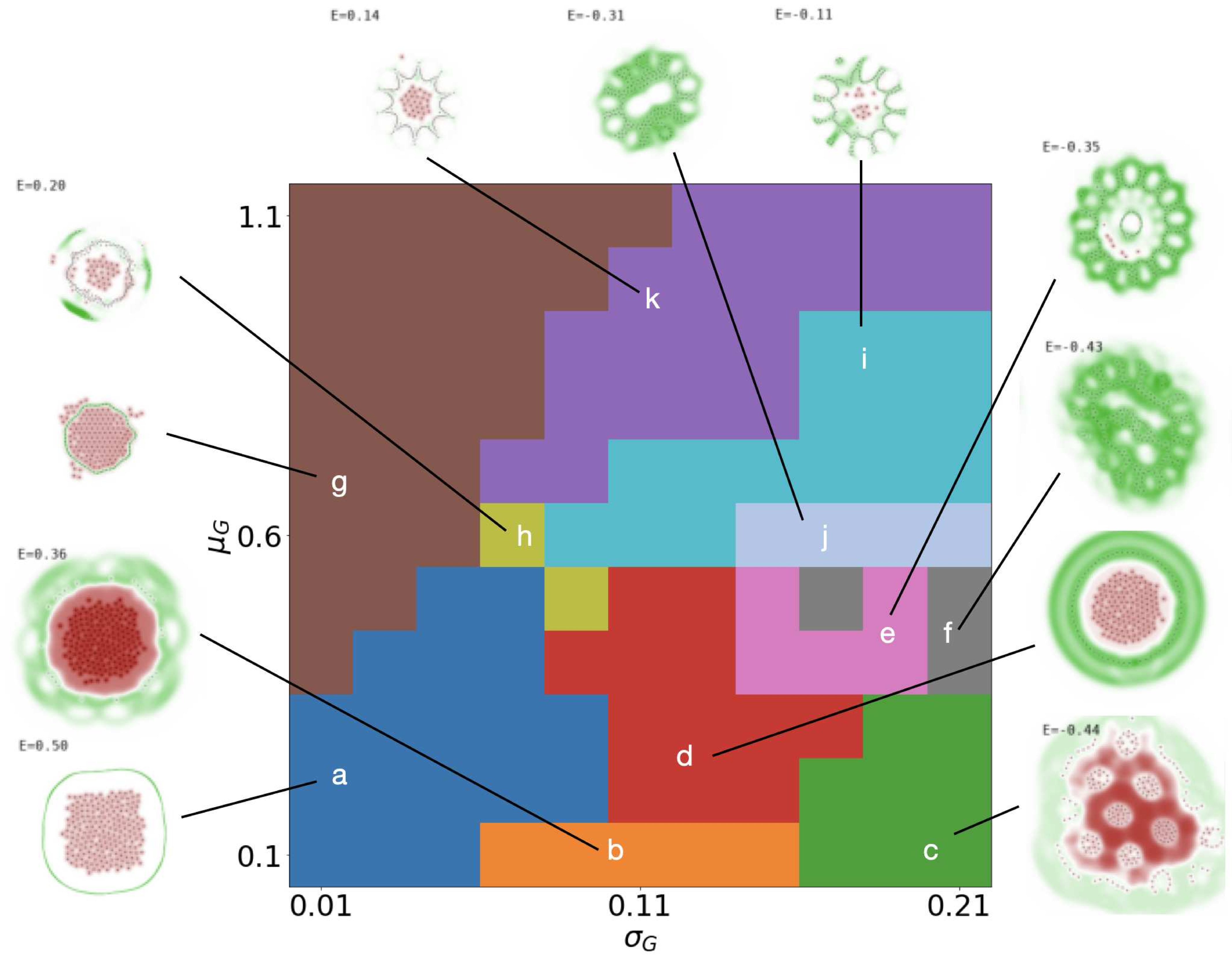}
      \caption{Phase diagram of stable configuration in pre--perturbation}
      \label{fig:fig2}
\end{figure}

\subsection{Diversity of patterns}
To investigate how stable configurations are distributed in the parameter space we calculated 100 trials for each of the $(\sigma_G, \mu_G)$ in the function $G$ of growth filed ${\bf G}^t$ in the range $0.01< \sigma_G < 0.21$ and $0.1< \mu_G <1.1$ around $(\sigma_G, \mu_G) = (0.15, 0,6)$, which Mordvintsev et al. have already found a stable pattern named ``rotor''(Figure \ref{fig:fig2}j).

\subsubsection{Stable configuration in pre--perturbation}

\begin{figure*}[htpb!]
      \centering
      \includegraphics[width=1\textwidth]{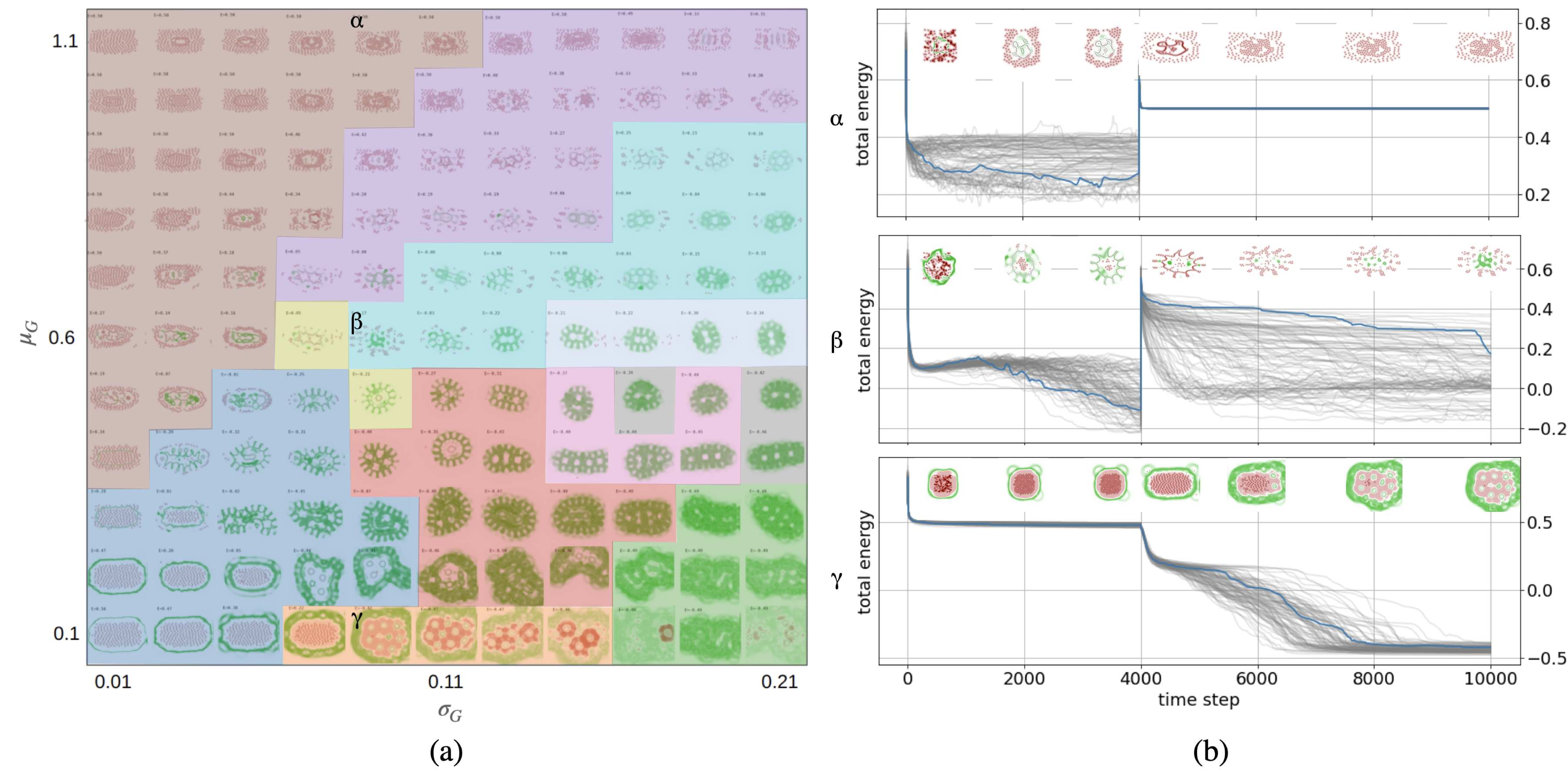}
      \caption{ Phase diagram of stable configuration in post--perturbation (a) Phase diagram: colors means pre--perturbation. (b) time evolution of the total energy. Parameters are $\mu_G = 1.1$, $\mu_G = 0.6$, and $\mu_G = 0.1$ for the upper, middle, and lower panels, respectively, and $\sigma_G = 0.09$ for all panels. Energy changes for 100 trials are shown in gray. The particle configuration at each time is shown from a randomly selected trial shown in the blue line.}
      \label{fig:fig3}
\end{figure*}

We found 11 different particle configurations that converge and maintain a constant value of the total energy from the initial condition pre--perturbation (Figure \ref{fig:fig2}). Although several stable particle configurations coexisted at the boundaries of different particle configurations, in the phase diagram, the particle configuration with the highest number of trials reached out of 100 trials was chosen as representative of that parameter set. The blue particle configuration is stable when both $\sigma_G$ and $ \mu_G$ were small. There was a cluster of high-energy particles in the center and a ring of low-energy particles a short distance around it (Figure \ref{fig:fig2}a, blue, later we call ``high dense core''). The orange one was similar to that of (a)``high dense core'', with high-energy particles in the center, but with a heterogeneous spread of lower-energy regions around it (Figure \ref{fig:fig2}b, orange, later we call ``solar flare''). The green one was similar to that of (a)``high dense core'' and (b)``solar flare'', with high-energy particles in the center, but the energy and particle arrangement were not homogeneous, forming multiple clusters. The high-energy region was surrounded by a broad, low-energy heterogeneous band (Figure \ref{fig:fig2}c, green, later we call ``pathological spots''). The red one was similar to that of (a)``high dense core'' and (b)``solar flare'', with a cluster of high-energy particles in the center, surrounded by a row of low-energy particles. In contrast to the (a)``high dense core'' and (b)``solar flare'', the low-energy region had two layers (Figure \ref{fig:fig2}d, red, later we call ``capsule''). The pink one had a ring of low-energy particles in the center, surrounded by sparse high-energy particles. Outside of the ring, a low-energy region was formed. This region was geometrically organized into compartments and looked like multiple spikes (Figure \ref{fig:fig2}e, pink, later we call ``sunflower''). The gray one was stable in energy, with the geometry of the (e)``sunflower'' broken, and there were no high-energy particles in the interior (Figure \ref{fig:fig2}f, gray, later we call ``fragile''). The brown one had a high-energy particle in the center, surrounded by a low-energy region. Higher energy particles also existed outside this energy (Figure \ref{fig:fig2}g, brown, later we call ``leaking core''). The yellow-green one had a high-energy particle in the center, surrounded by a band of high-energy particles. Outside of it were three separate low-energy regions (Figure \ref{fig:fig2}h, yellow-green, later we call ``three clouds''). The light blue one had some high-energy particles in the center, surrounded by low-energy particles. The surrounding particles were spine-like and had jagged outlines (Figure \ref{fig:fig2}i, light blue, later we call ``bugs''). The teal one had a particle-free region in the center, which was elliptical. Outside of that region were low-energy particles with multiple compartments. The particles were almost constant in energy, but the entire pattern was rotating, oscillating (Figure \ref{fig:fig2}j, teal, later we call ``rotor''). The purple one had a cluster of low-energy particles in the center and a zigzag shape on the outside, similar to ``bugs'' (Figure \ref{fig:fig2}k, purple, later we call ``sea urchin''). In summary, 11 different stable particle configurations were distributed in $\sigma_G$ and $\mu_G$ space.

To clarify how the parameters affect the transition of particle configurations, we compared the particle configurations when $\mu_G$ is fixed, and the value of $\sigma_G$ is increased. When $\mu_G = 0.1$ was fixed, and $\sigma_G$ was increased from small values, the stable particle configurations transitioned from (a) ``high dense core'' to (b) ``solar flare'' and then to (c) ``pathological spots'' (See Figure \ref{fig:fig2}). The outer low-energy ring spatially broadened and became a band, while the energy of the central group of particles increased. Multiple clusters were formed as $\sigma_G$ exceeded 0.17. When $\mu_G = 0.6$ was fixed, and $\sigma_G$ was increased from small values, the particle configuration shifted to (g) ``leaking core'', (h) ``three clouds'', (I) ``bugs'' and (j) ``rotor''. (g) ``leaking core'', where most particles were high energy in the center, turned into (h) ``three clouds'', where particles that had clustered in the center moved outward and formed low-energy regions on the outside. The outer low-energy region closed and formed a ring, growing in a zigzag structure, becoming a (j) ``rotor'', and finally, the group of high-energy particles in the center disappeared. When $\mu_G = 0.11$ was fixed, and $\sigma_G$ was increased from small values, a stable structure transitioned from a (g) ``leaking core'' to a (k) ``sea urchin''. In summary, the larger the value of $\mu_G$, the fewer the number of particles with high energy at the center of the particle configuration, and the larger the value of $\sigma_G$, the broader the spatial distribution of energy.

\subsubsection{Response configuration in post--perturbation}
To investigate the adaptivity of the stable particle configurations we found, we examined their response to a ``stretch'' perturbation at the 4,001st step and observed the subsequent time evolution over 6,000 steps for each set of parameters. For each parameter, we plotted the particle configuration at the end of the computation (the 10,000th step) for one randomly selected pattern from 100 independent trials (Figure \ref{fig:fig3}(a)).

To clarify how the parameters affect the particle configuration, we observed the time evolution of the total energy and the pattern change by varying $\mu_G$ at $\sigma_G = 0.09$. As shown in Figure \ref{fig:fig3}(b) $\gamma$ lower panel, for $\mu_G = 0.1$, the particle configuration was stable as (b) ``solar flare'' up to 4,000 steps (See also Figure \ref{fig:fig2}). After the perturbation, the particles stabilized gradually, forming a low-energy spot-like structure inside the higher-energy particle at the center. The final state was similar qualitatively to (c) ``pathological spots''. The total energy was around 0.5 in all trials before perturbation, and the total energy decreased with perturbation. The energy decreased gradually for each trial, and after 10,000 steps, it converged to a value near -0.5 for all 100 trials. As shown in Figure \ref{fig:fig3}(b) $\beta$ middle panel, for $\mu_G = 0.6$, the particle configuration was stable as (i) ``bugs'' up to 4,000 steps (see also Figure \ref{fig:fig2}). After the perturbation, the particles had high-energy particles on the outside and low-energy particles in the center. The low-energy particles moved irregularly, taking in outer particles, and the proportion of low-energy particles increased. The low-energy particles took a variety of shapes without having a regular shape. This characteristic of the particle arrangement was common regardless of the total energy value. The total energy decreased gradually for each trial trajectory of 100 trials from 2,000 steps to 4,000 steps. Immediately after the perturbation, the total energy increased immediately and then some trials dropped sharply after a few hundred steps, while some trials remained high until the end of the calculation, resulting in a wide range of final energy values, from -0.2 to 0.4. As shown in Figure \ref{fig:fig3}(b) upper panel $\alpha$, for $\mu_G = 1.1$, the arrangement of particles in the ring with lower energy was rapidly replaced while maintaining the (g)``leaking core'' particle arrangement up to 4000 steps (see also Figure \ref{fig:fig2}). After the perturbation, the low-energy ring disappeared, and at the same time, the structure of the inner particles that had been rapidly replacing their configuration was lost, after which all the particles were no longer in motion. Before the perturbation, the total energy fluctuated between 0.2 and 0.4 for each trial. After the perturbation, the total energy increased sharply for a brief period and quickly converged for nearly all trials to a value near 0.5. In summary, for a given value of the ratio of $\sigma_G$ to $\mu_G$, the range of possible values of the final total energy increases and multiple particle configurations become stable, a feature that is also observed for other $\sigma_G$.

\begin{figure}[ht]
      \begin{minipage}[b]{.95\hsize}
        \centering
        \includegraphics[width=1\textwidth]{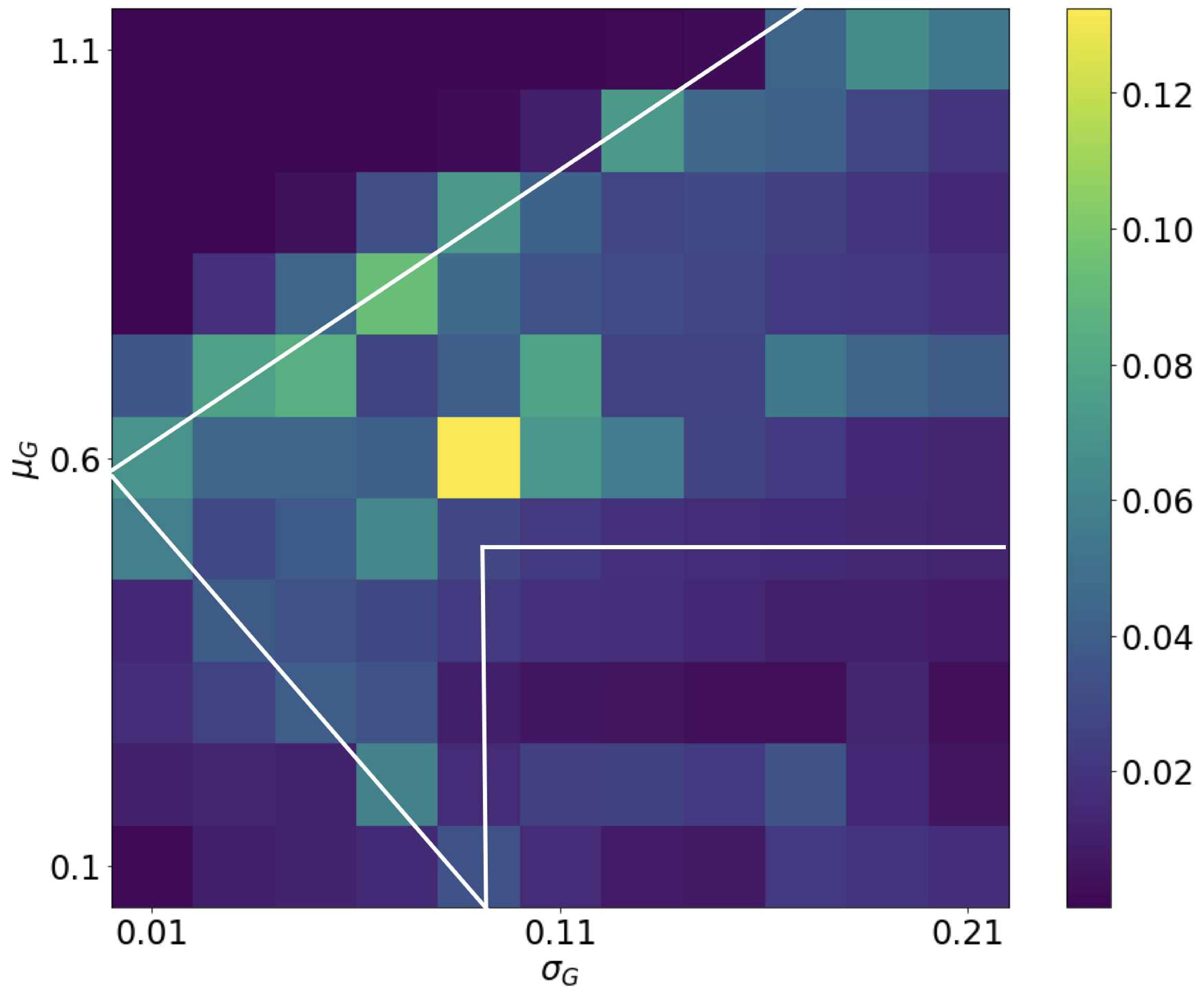}
        \subcaption{Diversity of metastable configuration}
      \end{minipage} 
      \hfill
      \begin{minipage}[b]{.95\hsize}
        \centering
        \includegraphics[width=1\textwidth]{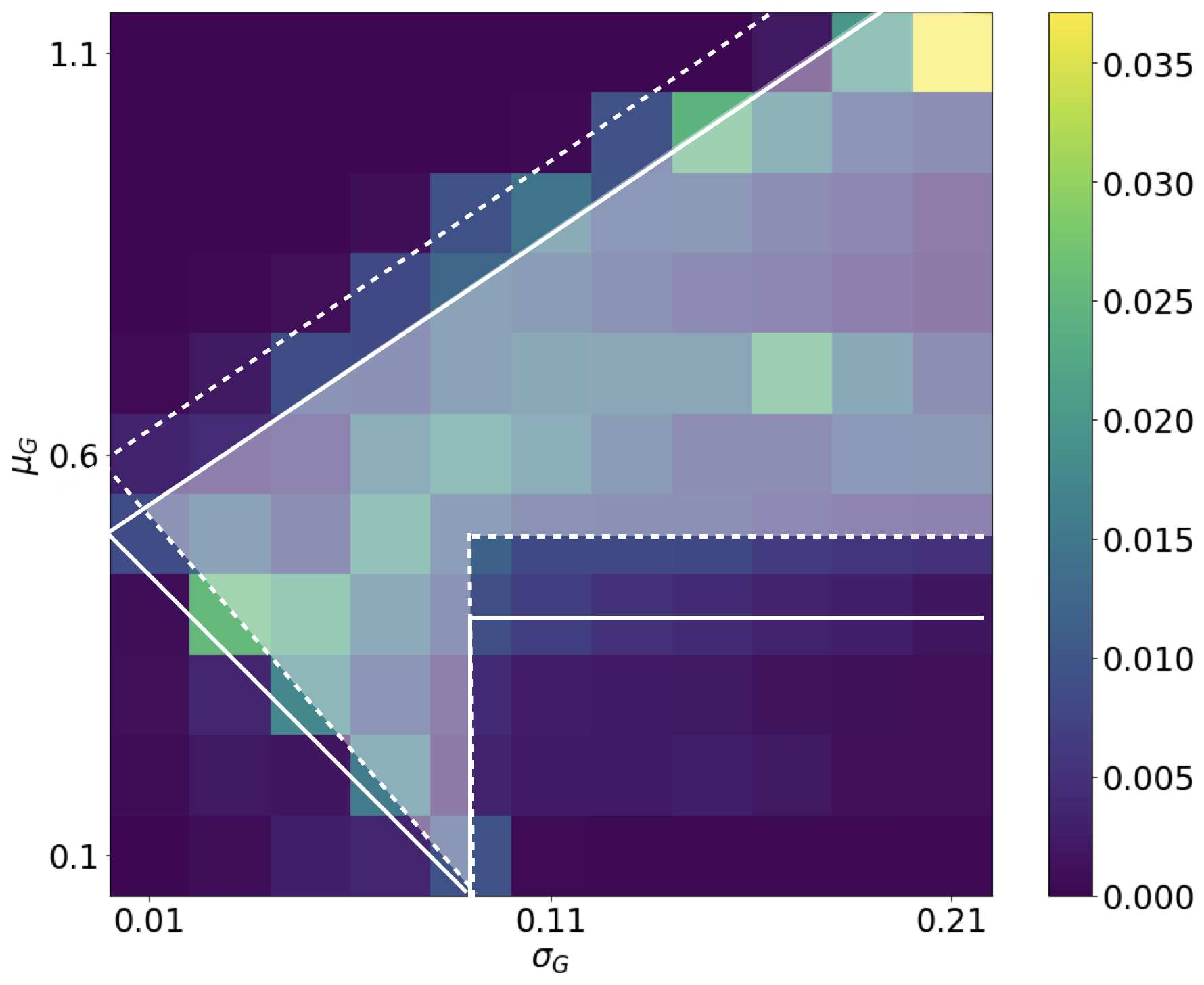}
        \subcaption{Time instability}
      \end{minipage} \\
      \caption{Evaluation of total energy stability for ``stretch'' perturbations}
      \label{fig:fig4}
\end{figure}

\begin{figure*}[htpb!]
    \centering
    \includegraphics[width=1\textwidth]{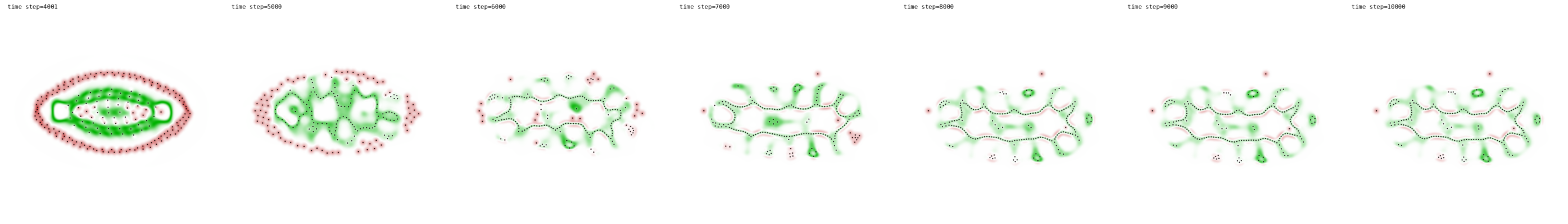}
    \caption{Example for dynamics of ``worm'' pattern in post--perturbation}
    \label{fig:fig5}
\end{figure*}

To quantitatively evaluate the dynamics of particle configurations, we measured adaptivity; the diversity of metastable configuration and time instability. The diversity of metastable configuration had a maximum at $\mu_G = 0.6$, and $\sigma_G = 0.09$. The parameter regions with relatively large diversity were localized in the region of $ - 25/4 * \sigma_G + 0.6 < \mu_G < 25/8 * \sigma_G + 0.6$ without $0.09 < \sigma_G < 0.21$ and $0.1 < \mu_G < 0.5 $ (Figure \ref{fig:fig4}(a), white line). In most of these localized parameter regions, the particle configuration was similar to ``worm'' (Figure \ref{fig:fig5}). After the perturbation, a small number of low-energy particles existed near the center of the particle cluster, surrounded by high-energy particles. The few low-energy particles migrated around and took in the high-energy particles around them, eventually taking in almost all of the particles and forming a spiky ``worm'' particle configuration (Figure \ref{fig:fig5}). The ``worms'' movement was irregular. This irregular movement affected the overall arrangement of the particles. The total energy in this particle configuration did not take a fixed value and was found to be stable over a wide range of values (Figure \ref{fig:fig3}, middle panel). In the rest of the parameter region, where $\mu_G$ is large and $\sigma_G $ is small, many particles remained high energy, did not change from their initial random configuration, and the total energy converged to a similar value in every trial (Figure \ref{fig:fig3}, upper panel). On the other hand, in the parameter region where $\mu_G$ is small and $\sigma_G$ is large, the particles interacted even with relatively distant particles, showing a pattern of sparsely distributed particles in space. Depending on the balance between $\sigma_G$ and $\mu_G$, particles self-organize like a dot pattern (Figure \ref{fig:fig3}, bottom panel). In summary, responding particle configurations can be classified into two types of configurations: those that stabilize and those that destabilize in terms of energy value. There were two types of stabilizing configurations: one in which particles remained high in energy without interacting with other particles, and one in which particles were sparsely distributed in space with low total energy.

The time instability had a maximum value at $\sigma_G = 0.21, \mu_G = 1.1$, and the high value region generally overlapped with the diversity of metastable configuration. Note that the parameter region on $ \mu_G = 25/4 * \sigma_G + 0.5 $ had low time instability, despite the high diversity of metastable states (Figure \ref{fig:fig4}(b)). The particle configurations observed in this parameter region had multiple metastable structures without ``worm'' appearing, and once stabilized in that stable pattern from the perturbation. In areas where both the diversity and the time instability are high (Figure \ref{fig:fig4}(b), highlight area), ``worm'' appear, the total energy does not take a constant value, has time instability, and by exploring the energy landscape, the total energy can be lowered could be done. In summary, for a ``stretch'' perturbation, a stable particle configuration transitions to another stable particle configuration. The manner of the transition is characterized by the time instability of the total energy and the diversity of metastable configuration, with ``worm'' particle configurations appearing in regions where both of the two measures take high values.

\section{Discussion}
\subsection{Summary of results}

In this study, we investigated the parameter dependency of the growth field $\bf G$ on the ``stretch'' perturbation of Particle Lenia, in order to validate its effectiveness for a novel model of autopoiesis. We found several stable configurations depending on the parameter. Upon applying a ``stretch'' perturbation to a stable particle configuration, a transition to another stable configuration occurred. The response to the perturbation was characterized by the diversity of metastable configuration and time instability. We found a ``worm'' configuration in which both of these two values are high. The ``worm'' was unstable in time but maintained its structure while dynamically changing its pattern and lowering its total energy. The conditions for the occurrence of this ``worm'' were determined by the ratio of $\mu_G$ to $\sigma_G$ and other parameters. The ability of the ``worm'' to maintain its structure while being temporally unstable to this environmental perturbation is considered to have adaptivity which is one of the autopoiesis's features.

\subsection{Future implications}
Although we examined the dependency of the growth factor $\bf G$ which determines the time evolution rule of the particle configuration, the role of the other parameters remains uninvestigated. Therefore, further exploration is necessary, specifically concerning the effect of newly introduced physical properties of particles, such as the strength and range of repulsion, on the particle dynamics, which is vital for assessing the expressive potential of Particle Lenia. Moreover, as Particle Lenia is an energy-based model, the particle dynamics can be construed as energy landscape trajectories. The used indicators, i.e., diversity of metastable configuration and time instability, likely represent the landscape's local minima count and complexity, respectively. By estimating statistical energy landscapes, such as Monte Carlo methods, we could analyze more detailed properties of the stability of the particle configurations.

Secondly, the diversity of metastable configuration and time instability used in this study as an indicator of adaptivity has not been correctly defined. The metastable state is defined in numerical calculations as the convergence of the total energy value to a certain value. A more accurate formulation of the metastable state is needed by drawing the energy terrain and measuring the response to perturbations.

Thirdly, We investigated the properties of the model with uniform parameters for all particles. It is worth analyzing the case where particles with different parameters are mixed. As shown in swarm chemistry \citep{sayama2009swarm}, particles with different parameters could emerge with a more complex and diverse particle configuration, so that we could find more various stable configurations.

Finally, We investigated the stability of particle configurations based on the temporal variation properties of the total energy of the system, however, to directly measure the agency of the particles other metric needs. Friston proposed to derive a Bayesian network from the physical configuration of the particles and its Markov blanket is the boundary that separates the agent and the environment \citep{friston2013life}. Although some arbitrariness in the derivation of the Markov blanket has been pointed out \cite{bruineberg2022emperor} and the derivation is frequently updated \citep{friston2021stochastic}, this boundary is applicable to particle systems, such as Particle Lenia. In addition, we can utilize other metrics, such as Bayesian network-based Mechanised causal graphs \citep{kenton2022discovering}, information theory-based indices of individuality \citep{krakauer2020information}, and Semantic information \citep{kolchinsky2018semantic}, and expect to be able to evaluate various aspects of agency.

\section{Conclusion}
Our paper broadens the scope of new applications of Particle Lenia as a tool for investigating autopoiesis. Although this work is only a partial parameter study, the capability of the model to generate complex particle dynamics brings us closer to implementing autonomous agents and robots with the properties of autopoiesis. Furthermore, considering human interaction with autonomous agents is an important process for realizing artificial life in the real world. Through further investigation of Particle Lenia and expansion of capacities, we expect that a greater variety of artificial agents will be developed.

\section{Data Sharing}
The code used to generate results is available as Colab notebook at \url{https://github.com/KazuyaHoribe/AdaptiveParticleLenia}. Some videos of pre- /post-perturbation also are shown in the same place.

\section{Acknowledgements}

This work was supported by JSPS KAKENHI Grant Number 23H04834.

We would like to thank Google Zurich team for sharing their codes.

\footnotesize
\bibliographystyle{apalike}
\bibliography{example} 

\end{CJK}
\end{document}